# Building a Smart, Secured and Sustainable Campus: A Self-Powered Wireless Network for Environmental Monitoring

Qutaiba I. Ai
University of Mosul
Iraq

**Background:** Traditional wired sensor networks for building environmental monitoring often face challenges related to scalability, cost, and resilience. This paper introduces a novel concept called the Green Environmental Monitoring Network (Green EMN), tailored specifically for campus environments.

**Objective:** The objective of this study is to propose a self-powered wireless network solution that utilizes strategically deployed wireless sensor nodes within buildings for environmental data collection, while integrating advanced security measures and sustainable power management strategies.

**Methods:** The Green EMN utilizes wireless sensor nodes deployed within buildings, transmitting data to gateway nodes equipped with the Embedded Cooperative Hybrid Intrusion Detection System (ECHIDS) for enhanced security. Sustainable power management techniques such as energy-efficient nodes, energy harvesting, and duty cycling are incorporated to ensure long-term operation.

**Results:** The proposed Green EMN offers significant advantages over traditional wired networks, including scalability, cost-effectiveness, resilience, and seamless integration with existing campus infrastructure. The innovative ECHIDS security system provides comprehensive threat protection against cyberattacks.

Conclusion:
By promoting a secure and sustainable monitoring environment, the Green EMN can greatly enhance environmental monitoring capabilities in campus buildings, fostering a healthier and more sustainable learning environment for students and faculty.

## I. INTRODUCTION

Creating and maintaining a healthy and sustainable environment within campus buildings is essential for the well-being of students, faculty, and staff. Environmental factors such as temperature, humidity, air quality, noise levels, and lighting significantly impact occupant comfort, health, and productivity. Therefore, continuous monitoring and effective management of these parameters are crucial for optimizing building operations and ensuring a positive learning and working environment [1,2].

Traditionally, wired sensor networks have been the primary tool for environmental monitoring in buildings. These systems offer reliable data collection and transmission but often suffer from several limitations [3,4]:

- Scalability: Wired networks are complex and expensive to install, especially in large or geographically dispersed campuses. Expanding or modifying the network can be challenging and costly.
- Cost: The installation and maintenance of wired infrastructure can be expensive, particularly in retrofitting existing buildings.
- Resilience: Wired networks are vulnerable to damage caused by physical infrastructure failures or natural disasters, leading to data loss and disruption of monitoring capabilities.
- Integration: Integrating wired networks with existing building management systems can be complex and require additional hardware and software.

The emergence of wireless sensor networks has revolutionized environmental monitoring by offering greater flexibility, scalability, and cost-effectiveness compared to wired systems. However, traditional wireless sensor networks often rely on battery-powered nodes, leading to concerns about [5,6]:

- Limited lifespan: Frequent battery replacements can be time-consuming, expensive, and generate environmental waste.
- Maintenance challenges: Manually replacing batteries in a large network of sensor nodes can be labor-intensive and disruptive.

Therefore, there is a growing need for sustainable wireless sensor network solutions that address these limitations and ensure long-term, reliable environmental monitoring in campus buildings.

The increasing demand for sustainable and efficient building operations has fueled the integration of Internet of Things (IoT) technologies into smart building ecosystems. This convergence unlocks exciting possibilities for optimizing building performance across various domains, including energy management, occupant comfort, and security. This review surveys the current landscape of research in this burgeoning field, highlighting key contributions and identifying promising avenues for future exploration.

**IoT Integration and Building Energy Management Systems (BEMS):** Several studies underscore the transformative potential of IoT in BEMS. [2] emphasizes the cost-effectiveness of IoT-based solutions, enabling ubiquitous monitoring and management of building energy consumption across domains like lighting, HVAC, and flexible loads. The authors present a compelling case study demonstrating how low-cost IoT devices can provide valuable insights into occupancy patterns and building usage, informing data-driven decision-making for improved energy efficiency. Similarly, the authors in [3] delve into the design and implementation of an IoT gateway for a cloud-based BEMS, focusing on its critical role in collecting and transmitting data from diverse devices and protocols, facilitating seamless integration within existing infrastructures.

**Fog Computing and Smart Building Platforms:** Fog computing emerges as a promising architectural paradigm for smart building applications. The authors in [4] propose a five-layer fog-based IoT platform, enabling distributed processing and real-time decision-making at the edge of the network. This architecture supports a diverse range of sensor and actuator nodes, empowering fine-grained monitoring of indoor ambience, occupancy, and lighting conditions. Reference [5] explores the application of IoT in a hybrid renewable energy system on a university campus, utilizing a four-layer architecture for data acquisition, communication, network modeling, and application development. This study showcases the versatility of IoT in integrating and optimizing complex energy systems within smart building environments.

**Challenges and Future Directions:** Despite significant advancements, several key challenges and future research directions warrant attention. Reference [6] provides a comprehensive review of IoT in smart buildings, outlining the major components, conventional architecture, and potential applications. The authors identify critical challenges like security, privacy, data management, feasibility, and collaboration between developers and the building industry. These challenges call for innovative solutions and collaborative efforts to ensure the seamless and secure integration of IoT technologies within smart building ecosystems. Similarly, the authors in [7] survey diverse smart building applications, highlighting areas like security control, energy management, HVAC monitoring, and healthcare for elderly residents. This broad overview underscores the vast potential of IoT to enhance various aspects of building operations and occupant well-being, prompting further research and development in specific application domains.

**Specific Applications and Techniques:** Several studies delve into specific applications and techniques within the smart building and IoT domain. Reference [8] proposes an IoT-based thermal modeling framework for learning building behavior using low-cost thermostats. This approach leverages machine learning to optimize energy consumption by dynamically adapting to occupant preferences and environmental conditions. The researchers in [9] present a cost-effective solution for non-smart residential appliances using smart load nodes, emphasizing its non-invasive integration within existing electrical infrastructures. This study demonstrates the potential of IoT to seamlessly retrofit existing buildings with intelligent control capabilities, contributing to improved energy efficiency without significant infrastructural modifications.

**IoT and Smart Grid Systems:** The convergence of IoT and smart grid systems holds immense potential for transforming energy management at the city and regional levels. Reference [10] provides a comprehensive survey of IoT-aided smart grid systems, covering architectures, applications, prototypes, challenges, and future directions. This study highlights the critical role of IoT in enabling real-time data collection, distributed control, and demand-side management within smart grids, paving the way for a more resilient and sustainable energy ecosystem. The authors of [11] review the architectures and functionalities of IoT-enabled smart energy grids, focusing on sensing, communication, and computing technologies. Security vulnerabilities, attack models, and mitigation techniques are also discussed. This emphasis on security is crucial for ensuring the reliability and integrity of interconnected smart building and grid systems.

**Energy Management and Microgrids:** The Authors in [12] review recent activities related to IoT-based energy systems, highlighting potential areas for improvement at different layers and reviewing communication technologies and standards. This comprehensive review provides valuable insights into the evolving landscape of energy management within smart buildings and microgrids, emphasizing the need for continued research and development in communication protocols, data interoperability, and control algorithms. Reference [13] proposes a hierarchical IoT-based microgrid framework for energy-aware buildings, showcasing its implementation in a laboratory setting. This practical demonstration underscores the feasibility and potential benefits of integrating microgrids with IoT-enabled building management systems, contributing to enhanced energy efficiency and grid resilience.

**Thermal Comfort and Occupancy-Driven Control:** Thermal comfort in buildings remains a crucial research area. The authors of [14] present a comprehensive review of this topic, identifying influencing factors, field surveys, measures for improvement, and energy-saving strategies. This review provides a valuable foundation for further research aimed at optimizing thermal comfort within smart buildings while minimizing energy consumption.

This paper proposes a new approach for environmental monitoring in campus buildings by adapting a self-powered Environmental Monitoring Network (EMN) concept. The proposed network addresses the limitations of traditional wired and battery-powered wireless systems by offering the following key features:

- Scalability and Flexibility: The wireless architecture enables easy deployment and expansion of the network across buildings and campuses.
- Cost-effectiveness: The use of energy-efficient nodes and sustainable power management strategies minimizes operational costs.
- Resilience: The self-powered design ensures continued operation even during power outages or infrastructure failures.
- Seamless Integration: The network can be easily integrated with existing building management systems for comprehensive data analysis and control.

The main contributions of this paper are as follows:

- We propose a novel architecture for a self-powered wireless network specifically designed for environmental monitoring in campus buildings.
- We present a comprehensive system design, including sensor selection, data communication protocols, power management strategies, and data visualization tools.
- We suggest the Embedded Cooperative Hybrid Intrusion Detection System (ECHIDS) for enhanced security. ECHIDS integrates diverse detection techniques, including signature-based, anomaly-based, and behavior-based methods, to provide comprehensive threat protection against various cyber attacks.
- We evaluate the performance of the proposed network through simulations and pilot deployments, demonstrating its effectiveness and scalability.

By addressing the limitations of existing solutions and offering a sustainable and scalable approach, this paper aims to advance the field of environmental monitoring in campus buildings, ultimately contributing to creating healthier and more sustainable learning environments.

## II. Materials and Methods

This section explores the key considerations and challenges involved in constructing a robust and environmentally sustainable EMN infrastructure. We delve into the critical aspects of self-sufficiency, power management, and network connectivity, while also introducing the concept of a "Green EMN Infrastructure" for minimal environmental impact.

To bolster the reliability of EMN communication systems, we propose a novel approach that utilizes wireless ad hoc network technology. Our strategy entails deploying a dynamic network infrastructure composed of diverse wireless nodes, encompassing both fixed and mobile units. Each node is specially designed to execute specific tasks aligned with the requirements of various EMN applications.

A central component of this network is the Wireless Solar Router (WSR). This multifunctional node plays a pivotal role in delivering a comprehensive range of essential EMN services to clients within its designated coverage area. These services include data and control signal transmission between critical and industrial facilities, text messaging capabilities, and even support for multimedia services. As crucial elements of the EMN infrastructure, WSRs receive data packets from various sources and subsequently relay them to a local or remote server located within a Monitoring and Control Center (MCC). Through collaborative efforts, these WSRs establish an ad hoc network, collectively transporting data packets to their designated destinations. This intricate network necessitates an efficient ad hoc routing protocol, as illustrated in Figure 1a.

Furthermore, each WSR functions as a router within the ad hoc network, enabling the forwarding of traffic originating from other WSRs towards their respective destinations. The adoption of ad hoc networking technology to enhance the reliability of EMN systems offers significant advantages over conventional wireless and wired methods. Ad hoc networks establish connections among nodes without relying on centralized infrastructure or administrative oversight, resulting in markedly reduced ownership, installation, and maintenance costs compared to other networking approaches.

Traditionally, WSRs have been positioned near wired electricity sources to ensure their power supply. However, this localized placement restricts the coverage area of the proposed EMN infrastructure and, consequently, the reach of its services. To overcome this limitation, we propose the implementation of self-powered WSRs. This innovative approach envisions WSRs harvesting and storing the energy they require from the surrounding environment, with a particular focus on solar energy, as depicted in Figure 1b. By harnessing renewable energy sources, WSRs can be deployed in virtually any location, irrespective of the availability of conventional power grids, significantly extending the coverage area of the EMN infrastructure and minimizing its environmental footprint.

The proposed "Green EMN Infrastructure" emphasizes sustainability and environmental consciousness throughout its design and operation. By leveraging self-powered WSRs and prioritizing renewable energy sources, we aim to minimize the network's dependence on fossil fuels and reduce its overall carbon footprint. This not only aligns with contemporary environmental goals but also ensures the long-term viability and resilience of the EMN infrastructure in the face of potential energy resource limitations.

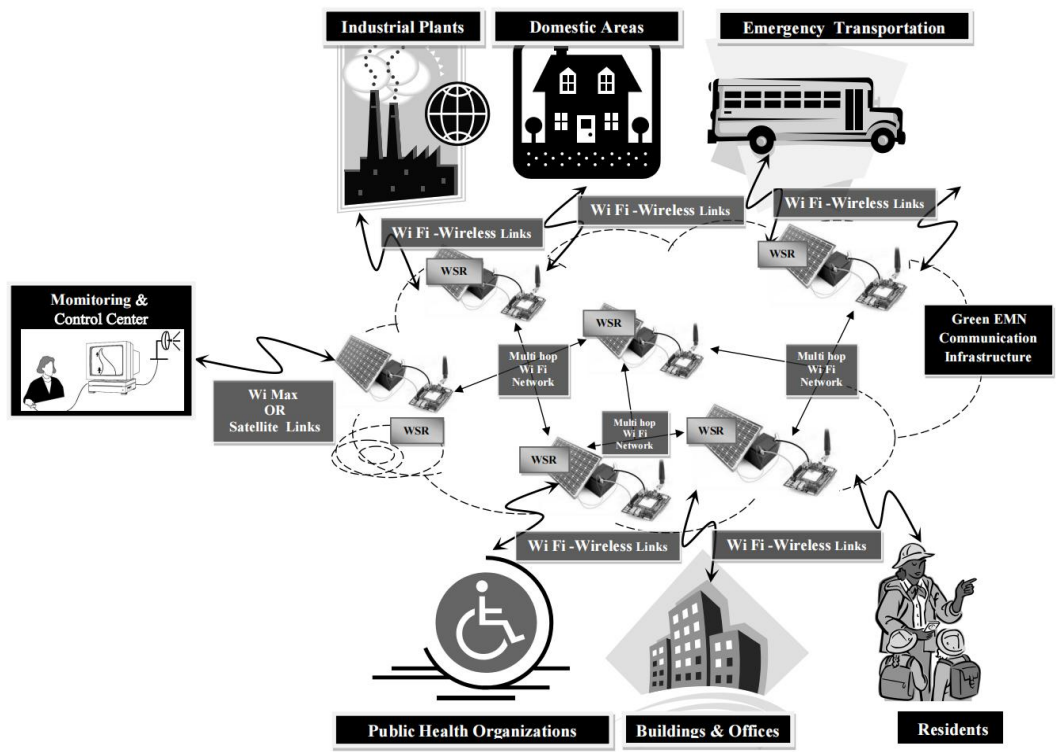

(a)

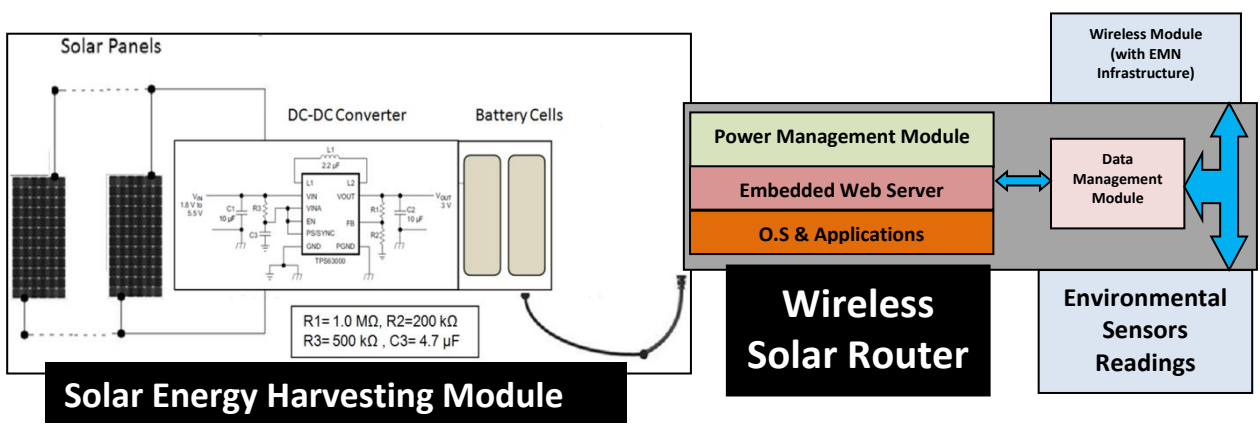

(b)
Figure(1): Self Powered EMN Infrastructure (a) EMN Topology (b) WSR Architecture

Next, we will unveil the "Duty Cycle Estimation (DCE) – Event Driven Duty Cycling (EDDC)" power management system, a cornerstone of our environmentally sustainable communication infrastructure. DCE-EDDC effectively harnesses solar energy and optimizes power consumption within the network, ensuring long-term operational viability and minimizing environmental footprint.

The versatile UBICOM IP2022 platform [15] serves as the foundation for our research, leveraging its adaptability and diverse functionalities. Due to the dynamic nature of network traffic, WSRs experience fluctuating power demands that significantly impact their lifespan and overall energy efficiency. Our primary objective is to equip WSRs with a reliable renewable energy source through solar panels (Figure 1b) and implement a sophisticated power management strategy to optimize battery utilization and minimize energy expenditure.

The solar energy harvesting module incorporates a dedicated circuit meticulously designed to collect solar energy, manage battery storage, and deliver power to the WSR. We employ the Texas Instruments TPS63000 low-power boost-buck DC-DC converter [16] to ensure consistent and reliable voltage supply to the embedded system.

To further enhance energy efficiency, we introduce a novel duty cycling methodology, "DCE-EDDC," which involves strategically transitioning between active and sleep states to conserve power. During active states, WSRs execute their core functions, while sleep states minimize energy expenditure by restricting operations to essential tasks [17, 18].

Our proposed DCE algorithm dynamically adjusts WSR operations based on available energy, establishing a direct correlation between service rate and power budget. Sleep periods are calculated daily, considering factors such as power consumption, weather conditions, and stored energy. This dynamic approach leverages the well-defined relationship between duty cycling periods, Average Service Rate (ASR), and Available Energy (AE):

- **Average Service Rate (ASR):** The average data traffic (bps) transmitted and received by the WSR.
- **Duty Cycling Periods:** Time is divided into slots, and the duty cycle represents the ratio of sleep periods to the total slot time.
- **Available Energy (AE):** The sum of residual battery energy from the previous day and anticipated solar energy for the next day.

As illustrated in Figure 2a, each time slot is divided into active and sleep periods. The WSR initially enters a predetermined sleep duration calculated based on AE. Upon entering the active period, the WSR awakens and processes stored packets in the WLAN NIC.

EDDC governs WSR behavior during active periods, considering both scheduled tasks and responses to incoming packets. This technique leverages "Clock Stop Mode," where the system clock and CPU core clock are disabled, putting the WSR motherboard into a low-power state. While the clock is inactive, the interrupt logic and sleep timer remain operational. The WSR can transition out of this sleep mode through sleep timer interrupts or external interrupts from the WLAN NIC. Crucially, program execution resumes from the paused point without a chip reset. This mode effectively shifts the motherboard to a power-saving state while maintaining essential functions like external interrupt circuits, the sleep timer, and program memory. Upon receiving an interrupt, such as a packet reception or sleep timer expiration, the board promptly awakens and executes necessary actions within a few clock cycles (Figure 2b).

By combining DCE's dynamic sleep period estimation with EDDC's efficient event-driven wake-up mechanism, our power management system significantly reduces energy consumption while ensuring reliable data transmission and service delivery within the EMN infrastructure. This innovative approach contributes to the creation of a sustainable and environmentally conscious communication network, paving the way for a greener future.

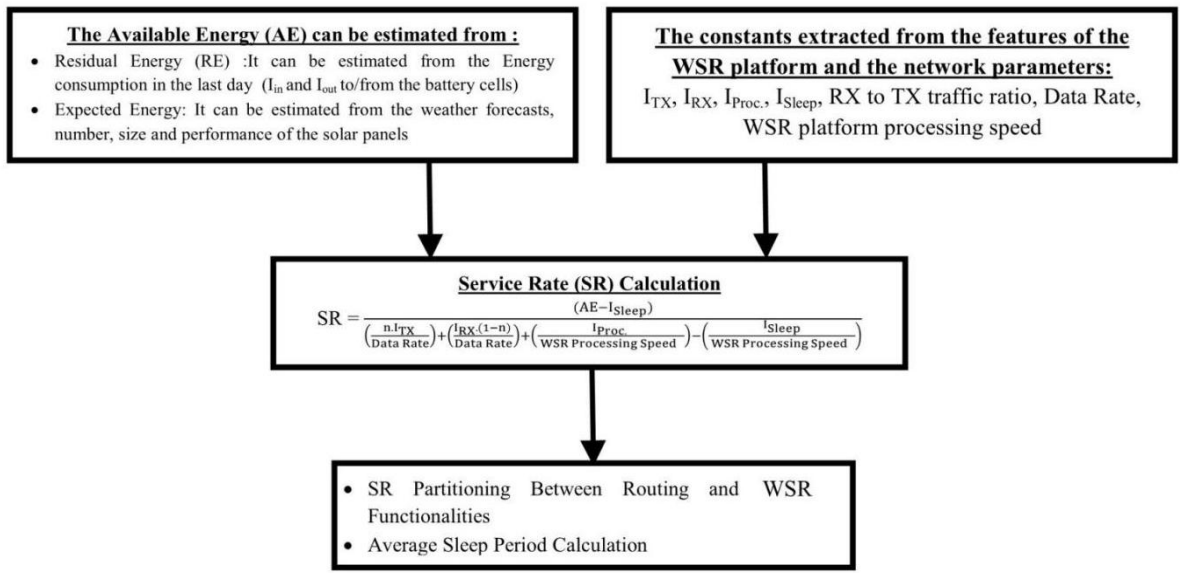

(a)

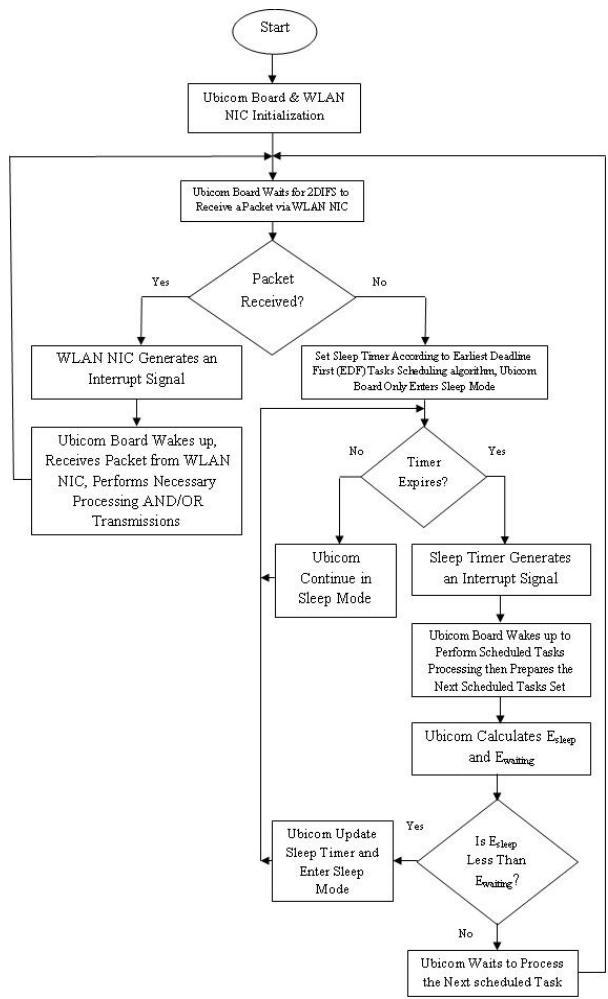

(b)

Figure(2): DCE-EDDC Algorithm (a) DCE Algorithm (b) Flowchart of EDDC Algorithm

### III. Results

To rigorously validate the proposed Green EMN infrastructure's performance, practicality, power efficiency, and resilience, we outline a comprehensive experimental platform for thorough evaluation. This section delves into the key metrics employed to assess various aspects of the system's functionality under diverse operating conditions.

To accurately gauge the power consumption of the Ubicom WSR under realistic EMN traffic scenarios, we propose a dedicated experimental framework depicted in Figure 3. This framework involves the creation of diverse EMN traffic profiles that closely resemble real-world network conditions. These profiles are then fed into an OPNET IT Guru simulation model, which serves as a representative emulation of the EMN infrastructure and facilitates the generation of the necessary network traffic. Subsequently, a traffic generator PC utilizes this simulated network traffic to mimic the behavior of the EMN in interaction with a WSR. By meticulously monitoring the WSR's power consumption during these simulated interactions, we can obtain valuable insights into its energy efficiency under varying traffic loads and network conditions.

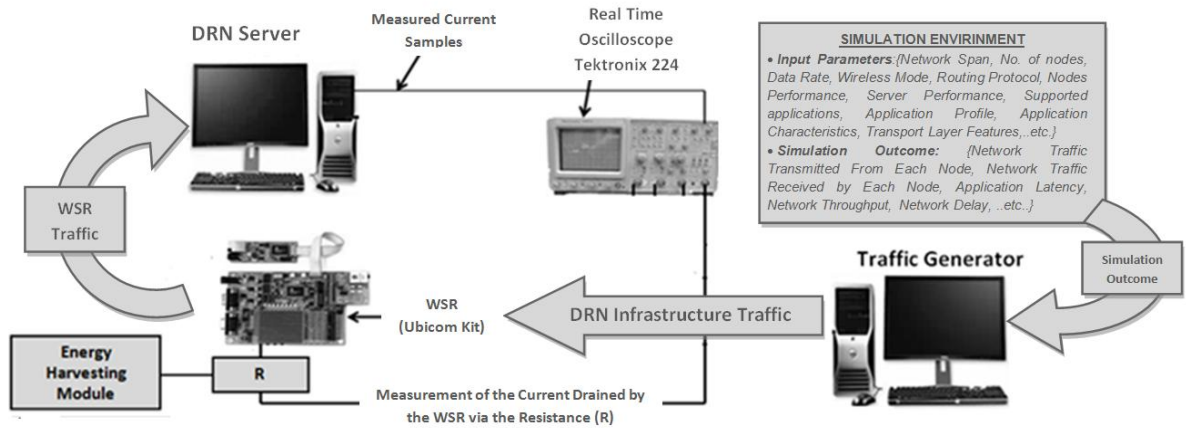

Figure(3): The Experimental Framework

Our evaluation commences with the selection of a real-world map encompassing an area of 5x5 square kilometers, serving as a representative environment for deploying 40 WSRs within the simulated EMN infrastructure [15]. Prior analysis [18] identified Optimized Link State Routing (OLSR) as the most effective protocol for dynamic ad hoc topologies, making it the cornerstone of our simulation model. To faithfully integrate OLSR's functionalities, we incorporated a predefined set of parameters specified in OLSR RFC 3626 [15], as detailed in Table 1.

Furthermore, to comprehensively assess the network's performance, we introduce various EMN services into the simulation environment, as outlined in Table 1. These services will generate realistic network traffic patterns, enabling us to rigorously evaluate the system's ability to handle diverse data transmission demands.

By meticulously designing these simulation parameters and incorporating representative EMN services, we establish a robust foundation for evaluating the proposed Green EMN infrastructure's performance, power efficiency, practicality, and resilience under various operating conditions.

Table(1): Simulation Model Parameters

| | |
|---|---|
| **Simulation Time (Minute)** | 60 |
| **No. of WSR nodes** | 40 |
| **Network Span Area (Km$^2$)** | 25 |
| **WSR Modeling Parameters** | Packets Processing Rate (Packet/sec.) = 2000<br>Memory = 2 M Byte |
| **WLAN settings** | Data Rate (Mbps) : 18 for IEEE802.11a |

| **OLSR settings** | | Hello Interval(sec.) = 2<br>TC Interval(sec.) = 5<br>Neighbor Hold Time(sec.) = 6<br>Topology Hold Time(sec.) = 15<br>Duplicate Message Hold Time(sec.) = 30 |
|---|---|---|
| **EMN Applications** | EMN Signaling | 3 Sensors/WSR<br>Sensors to WSR packet length = (128-512) Bit<br>Sensors to WSR packets rate = (20-1000) Packet/s |
| | Text Messaging | Message Inter-arrival Time = 10 s<br>Message Size = (200 to 1000)Byte [Uniformly Distributed] |
| | Image File Transfer | 1024×768 Pixels (JPEG Compression)<br>File Size = (0.5-1) MByte<br>Inter-request Interval = 180 s [Poisson Distribution] |
| | Multimedia Streaming | 352 × 288 @ 15 fps<br>197–421 Kb/s<br>H.264/AVC |

Our forthcoming experiments delve into two key metrics to quantify the effectiveness of the DCE-EDDC power management system:

1. Normal Mode Current Consumption: This baseline measurement assesses the power draw of the Ubicom board and its peripherals without any power management strategies in place.

2. Sleep Mode Current Consumption: This mode showcases the impact of DCE-EDDC, meticulously monitoring the power consumption of the Ubicom board and its accessories under the implemented power management scheme.

To demonstrably illustrate the advantages of DCE-EDDC, we will conduct a series of tests leveraging the experimental setup depicted in Figure 3. Our primary investigative focus will be to comprehend how varying the number of EMN clients served by each WSR influences network traffic, as illustrated in Figure 4. We will meticulously analyze various TCP/IP protocols, with a particular emphasis on application layer traffic flowing in both directions, as these represent the major contributors to network load. Other traffic sources, such as layer 2 and OLSR-related traffic, will have a comparably smaller impact. Notably, the WSR's location within the network topology significantly influences its traffic volume, with higher traffic (and consequently, higher power consumption) observed in WSRs closer to the EMN server. These strategically chosen WSRs represent the most power-intensive scenario within our experimental model.

Our assessment will also delve into the impact of the DCE algorithm under diverse scenarios, with initial settings outlined in Table 2. These experiments aim to evaluate the algorithm's adaptability to various operational conditions, considering fluctuations in Available Energy (AE) levels. Table 3 presents corresponding values of Average Service Rate (ASR) and Average Sleep Period (ASP) resulting from these scenarios, with Residual Energy (RE) representing battery charging levels and (N) indicating the number of paralleled solar panels. The DCE algorithm's ability to dynamically adjust duty cycling based on available energy levels ensures that the WSRs continue to operate in a pre-planned and managed manner.

Furthermore, we aim to evaluate the effectiveness of DCE-EDDC in safeguarding against unmanaged network traffic conditions, such as those arising from an Energy Exhaustive Denial-of-Service (DoS) Attack [15]. We will direct varying network traffic rates at the WSR, both with and without the implementation of the proposed power management method. Figure 5 will provide insights into how well the managed WSR sustains its battery life, irrespective of fluctuations in incoming traffic rates. In contrast, the unmanaged WSR's power consumption significantly increases with varying traffic rates, leading to rapid battery depletion.

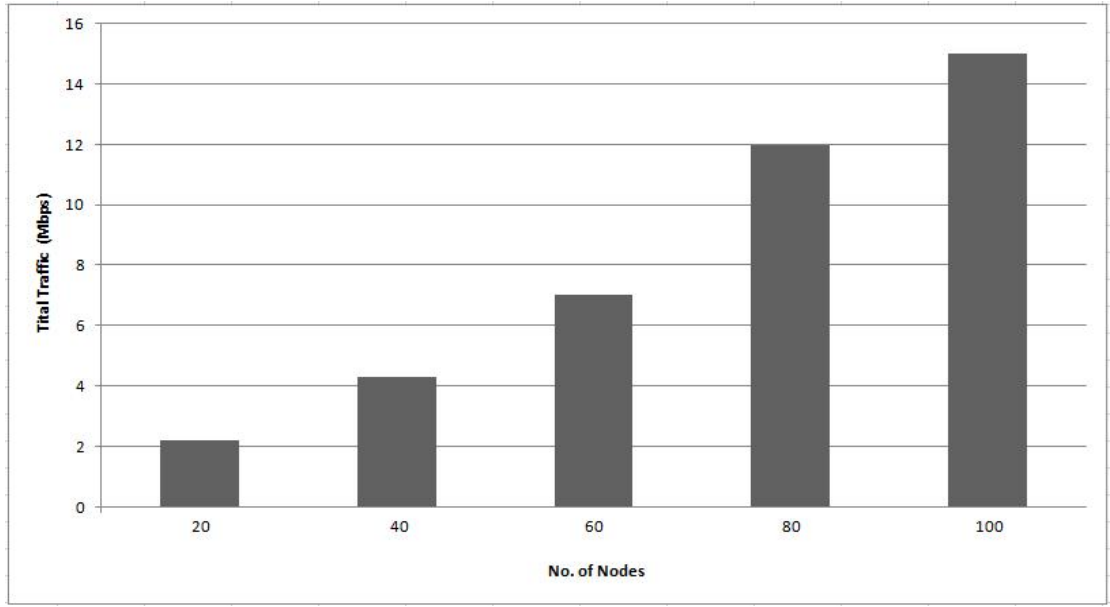

Figure(4): WSR Traffic Statistics

Table (2): Initial settings of DCE experiments

| Data Rate | 18 Mbps (IEEE802.11a) |
|---|---|
| Traffic Characteristics | $n = (\frac{RX\ Traffic}{TX\ Traffic}) = 4$ |
| Data Processing Speed of the WSR | 24 Mbps |
| $I_{TX}$ | 150 mA |
| $I_{RX}$ | 120 mA |
| $I_{Proc.}$ | 150 mA |
| $I_{Sleep}$ | 1 mA |
| Battery Characteristics | 3 v, 2800mAh |
| Solar Panel | SP1 |
| Average Current Produced in a Sunny Day | 34 mA |
| Effective Charging Time | 15 Hours |
| Average Current Produced in a Cloudy Day | 20.5 mA |
| Effective Charging Time | 13 Hours |
| Average Current Produced in a Rainy Day | 14.4 mA |
| Effective Charging Time | 11 Hours |

Table (3): ASR & ASP Values Under Different Conditions

| RE (% of Battery Capacity) | N | Weather Condition | AE (mAh) | ASR (Mbps) | ASP (s) |
|---|---|---|---|---|---|
| 100% | 1 | Sunny | 4960 | 7.75 | 0.57 |
| 100% | 1 | Cloudy | 3801 | 5.94 | 0.67 |
| 100% | 1 | Rainy | 3482 | 5.44 | 0.70 |
| 75% | 1 | Sunny | 4260 | 6.66 | 0.63 |
| 75% | 1 | Cloudy | 3101 | 4.85 | 0.73 |
| 75% | 1 | Rainy | 2782 | 4.35 | 0.76 |
| 50% | 1 | Sunny | 3560 | 5.56 | 0.69 |
| 50% | 1 | Cloudy | 2401 | 3.75 | 0.79 |
| 50% | 1 | Rainy | 2082 | 3.25 | 0.82 |
| 25% | 1 | Sunny | 2860 | 4.47 | 0.75 |
| 25% | 1 | Cloudy | 1701 | 2.66 | 0.85 |
| 25% | 1 | Rainy | 1382 | 2.16 | 0.88 |
| 100% | 2 | Sunny | 7120 | 11.13 | 0.38 |
| 100% | 2 | Cloudy | 4802 | 7.50 | 0.58 |
| 100% | 2 | Rainy | 4164 | 6.51 | 0.64 |
| 75% | 2 | Sunny | 6420 | 10.03 | 0.44 |
| 75% | 2 | Cloudy | 4102 | 6.41 | 0.64 |
| 75% | 2 | Rainy | 3464 | 5.41 | 0.70 |
| 50% | 2 | Sunny | 5720 | 8.94 | 0.50 |
| 50% | 2 | Cloudy | 3402 | 5.32 | 0.70 |
| 50% | 2 | Rainy | 2764 | 4.32 | 0.76 |
| 25% | 2 | Sunny | 5020 | 7.84 | 0.56 |
| 25% | 2 | Cloudy | 2702 | 4.22 | 0.77 |
| 25% | 2 | Rainy | 2064 | 3.23 | 0.82 |

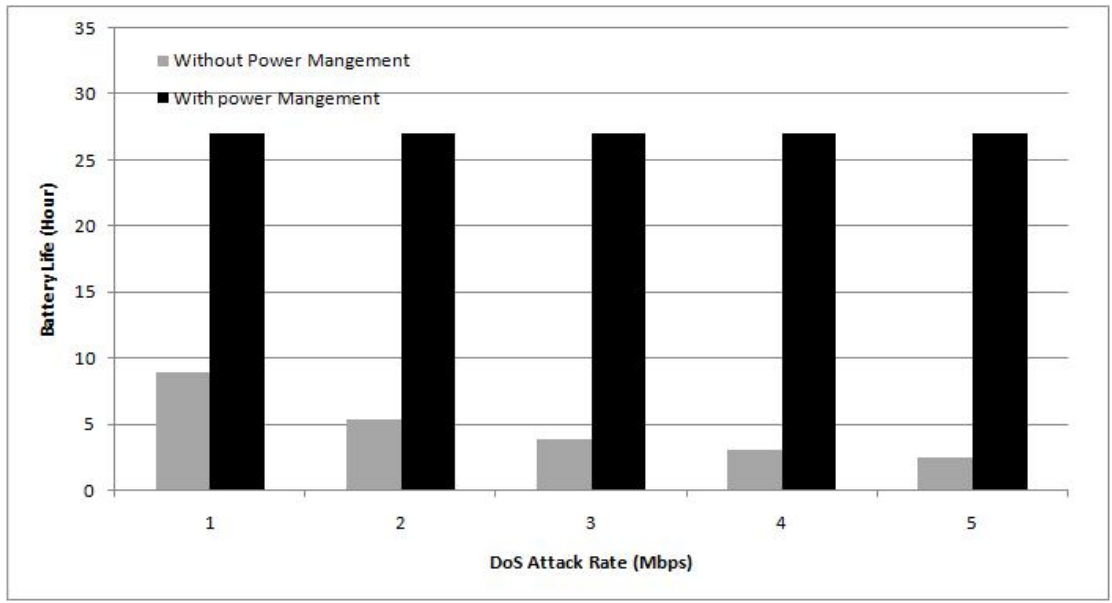

Figure(5): WSR Battery life according to different DoS attack rates

To ensure the availability and integrity of the Green EMN infrastructure, we propose a Cooperative Hybrid Intrusion Detection System (ECHIDS) that leverages the combined strengths of signature-based, anomaly-based, and behavior-based detection techniques. This collaborative approach empowers WSRs to work together with the MCC server, exchanging security data to gain a comprehensive understanding of the system's security posture, potential threats, and their origins.

- WSRs as IDS Sensors: WSRs act as distributed IDS sensors, periodically sending security reports to the MCC server. These reports include details about the number, types, and causes of attacks detected on each WSR.
- MCC Server as Central Hub: The EMN server aggregates the reports, analyzes them to generate a comprehensive security overview, and prescribes appropriate countermeasures. It then communicates these countermeasures to both the WSRs and the EMN administrator.

In this context, we are leveraging diverse detection techniques:

- Signature-based IDS: This component, based on the open-source SNORT IDS, resides on WSRs and is equipped with predefined rulesets targeting known threats. The EMN server aggregates and analyzes reports from WSRs to identify the most prevalent threats and dynamically updates the SNORT rulesets on WSRs to enhance detection efficiency.

- Anomaly-based IDS: This component utilizes an Artificial Neural Network (ANN) to detect deviations from normal system behavior. It analyzes overall network activity (e.g., data rates) on various timeframes (days, months, years) to identify anomalies indicative of potential attacks.
- Behavior-based IDS: This component focuses on identifying EMN - specific attacks like Black Hole and Energy Exhaustive attacks by monitoring activities against pre-defined behavioral rules.

The core of ECHIDS lies in its Hybrid Intrusion Detection System (HIDS), which integrates all three detection methods within each WSR (see Figure 6). Incoming traffic is first preprocessed to extract key features like average and peak data rates, and burst size. This preprocessed data is then fed to the three individual IDSs, each contributing to the final verdict. The final decision triggers actions such as generating security reports, blocking malicious nodes, and updating routing tables to avoid compromised paths. Benefits of ECHIDS can be listed below:

- Comprehensive Threat Detection: ECHIDS combines the strengths of multiple detection techniques to provide a wider and more accurate view of the threat landscape.
- Collaborative Security: Distributed threat detection and centralized analysis enable coordinated responses to attacks across the entire EMN infrastructure.
- Resource Efficiency: By leveraging the DRN server for rule updates and pre-processing, ECHIDS minimizes resource consumption on individual WSRs.

By deploying ECHIDS, the Green EMN infrastructure gains a robust and adaptable security system, capable of effectively safeguarding against diverse threats and ensuring the reliable and secure operation of the network.

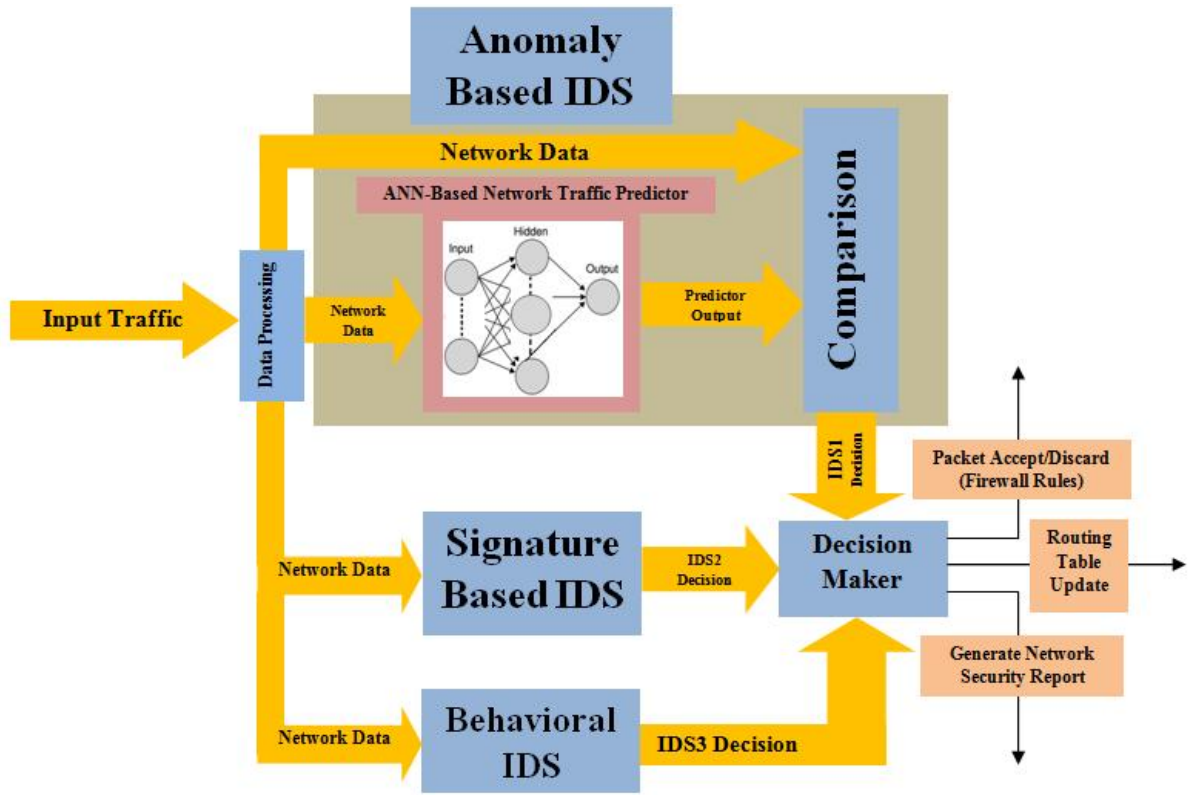

Figure(6): ECHIDS Architecture

## IV. Discussion

Table 4 summarizes the key statistical findings gleaned from our comprehensive experiments, meticulously examining the combined impact of the proposed power management techniques and the EMN algorithms. The compelling data presented in this table reveals a significant reduction in the number of solar panels required and a decrease in the necessary battery capacity for WSRs operating in Sleep mode to support various EMN applications. These remarkable results demonstrably showcase the effectiveness of the proposed power management scheme in extending the operational lifespan and enhancing the overall reliability of solar-powered, battery-backed WSRs. This ultimately translates to a positive contribution towards establishing a more robust and sustainable EMN infrastructure.

Furthermore, it is noteworthy that the network's performance, installation time, and cost all adhere to the predefined requirements and remain within the acceptable range for typical EMN operations, as documented in relevant studies [10-12]. This comprehensive analysis underscores the practical feasibility and cost-effectiveness of the proposed Green EMN infrastructure with DCE-EDDC, solidifying its potential for real-world environmental monitoring and control applications.

Table(4) Evaluation Metrics of Green EMN Infrastructure

| Network Type | | Wireless Mesh Network (WMN) |
|---|---|---|
| **Ad Hoc Routing Protocol** | | Optimized Link State Routing (OLSR) |
| **WSR Node** | | Ubicom IP2022 Network Processor Platform, 120 MHZ CPU, 2 Mbyte memory |
| **Wi Fi Standard** | | IEEE802.11a, 18 MHZ |
| **WSR Transmit Power (W)** | | 25dBm |
| **Antenna Type** | | Omni Antenna 2.4GHz/5GHz Dual Band Gain: 1.5 dBi (2.4GHZ) / 4.5 dBi (5GHZ) |
| **WSR Radial Transmit Range (m)** | | 300 |
| **Average Installation Time/WSR** | | 30 Minute |
| **Estimated Cost/WSR** | | 150 $ |
| **Quality of Service (QoS) Support** | | Yes |
| **Remote WSR Management** | | Yes, via Simple Network Management Protocol (SNMP) |
| **Network Performance** | Average Network Access Delay (s) | 0.00169 |
| | Average File Transfer Latency (s) | 6 |
| | Average Video Streaming Latency (s) | 0.5 |
| **Requirements of Solar System** | Solar Panel Dimensions (High (cm) × Width (cm)) | 36 × 29 |
| | Mean Voltage (v) | 4 |
| | Mean Current (mA) | 360 |
| | Rated Power (w) | 1.5 |
| | No. of Paralleled Solar Panels Required [Normal Mode] | 6 |
| | No. of Paralleled Solar Panels Required [Sleep Mode] | 2 |
| | Battery Capacity (mAh) [Normal Mode] | 2300 |
| | Battery Capacity (mAh) [Sleep Mode] | 750 |
| **Power Consumption Analysis** | Average Drained Current (mA) [Normal Mode] | 150 |
| | Average Drained Current (mA) [Sleep Mode] | 70 |
| | Battery Life (Hour)/Normal Mode [2800 mAh Battery Cells] | 15 |
| | Battery Life (Hour)/Sleep Mode [2800 mAh Battery Cells] | 29 |

To comprehensively assess the performance and resource utilization of the proposed ECHIDS, we employed established evaluation metrics outlined in [15]. Our implementation comprised 1000 SNORT rules, an anomaly-based IDS with a weekly traffic predictor, and a behavior-based IDS designed to detect Black Hole and Energy Exhaustive DoS attacks.

Table (5) summarizes the evaluation of ECHIDS. Key insights gleaned from the analysis include:
- Enhanced Security and Management: ECHIDS offers detection capabilities for a diverse range of attack patterns through its extensive pool of SNORT rules. Additionally, its flexibility allows for potential future enhancements to address intricate ad hoc network attacks.
- Remote Management Capabilities: Notably, the UBICOM platform incorporates a pre-configured SNMP agent, which plays a vital role in enabling remote management and reconfiguration of both WSRs and ECHIDS, streamlining network maintenance and security updates.

In conclusion, the evaluation results demonstrate that ECHIDS effectively delivers comprehensive security features while maintaining efficient resource utilization and minimal performance impact within the WSR platform. This combination of robust security, resource optimization, and remote management capabilities underscores the value of ECHIDS for safeguarding the Green EMN infrastructure and ensuring its reliable and secure operation.

Table(5) Evaluation Metrics of the Suggested ECHIDS

| ECHIDS Security Feature | Depth of System's Detection Capability | (+3500 SNORT Attack Signatures + Ad hoc networks behavioral attacks) |
|---|---|---|
| | Firewall Interaction | Supported & Integrated |
| | Router Interaction | Supported & Integrated |
| | Simple Network Management Protocol (SNMP) Interaction | Supported & Integrated |
| | Hybrid Intrusion Detection System | Supported & Integrated |
| SECURITY ASSESSMENT OF ECHIDS | | |
| Attack Type | Attacks' Target | Defense Strategy |
| Known Internet Attack Patterns (more than 3500 as defined by SNORT IDS developers) | • WSRs' hardware, software, services and energy resources<br>• Functionality of network, transport and application layers<br>• Network Infrastructure | Embedded SNORT (Signature) Based IDS |
| Denial of Service Attacks | • WSRs' hardware, software, services, communication and energy resources<br>• DRN Infrastructure | • Embedded SNORT (Signature) Based IDS<br>• Anomaly Based IDS |
| Distributed Denial of Service Attacks | • WSRs' hardware, software, services, communication and energy resources<br>• EMN Infrastructure | • Embedded SNORT (Signature) Based IDS<br>• Anomaly Based IDS |
| AD Hoc Routing Attacks (e.g., Black hole and Worm hole attacks, etc..) | AD Hoc Routing Protocols | Behavioral Based IDS |
| Sybil attack | WSR services | • Behavioral Based IDS<br>• Entity Authentication |
| Timing Attack | WSR services | • Embedded SNORT (Signature) Based IDS<br>• Behavioral Based IDS |
| Energy Exhaustive attack | WSR energy resources | • Behavioral Based IDS<br>• Anomaly Based IDS |
| Application Attack | WSR services | • Embedded SNORT (Signature) Based IDS<br>• Behavioral Based IDS |

## V. CONCLUSION

This paper has presented a comprehensive vision for a novel Green Environmental Monitoring Network (Green EMN) infrastructure, encompassing innovative features to deliver a self-powered, reliable, and secure solution for diverse environmental monitoring applications. Our design incorporates several key advancements, each contributing to the overall effectiveness and sustainability of the system:

• Intelligent Power Management with DCE-EDDC: The core of the Green EMN lies in the "Duty Cycle Estimation (DCE) – Event-Driven Duty Cycling (EDDC)" power management algorithm. This method excels in its simplicity, effectiveness, and adaptability, optimizing energy resource utilization and network performance across various operational scenarios. By dynamically adjusting duty cycles based on available energy and application demands, DCE-EDDC significantly reduces the number of solar panels and battery capacity required, translating to substantial cost savings and environmental benefits.

- Embedded Cooperative Hybrid Intrusion Detection System (ECHIDS): To ensure the security and integrity of the collected data, we propose ECHIDS, a novel intrusion detection system integrated within each Wireless Solar Router (WSR). ECHIDS leverages the combined strengths of signature-based, anomaly-based, and behavior-based detection techniques, providing comprehensive protection against a wide range of cyberattacks. This collaborative approach enables distributed threat detection and centralized analysis, safeguarding the network against both known and emerging threats.
- Scalable and Sustainable Architecture: The Green EMN architecture is designed for scalability and ease of deployment. Wireless sensor nodes offer flexibility in placement and adaptation to diverse monitoring requirements, while strategically placed gateway nodes facilitate efficient data aggregation and communication. By seamlessly integrating with existing campus infrastructure, the Green EMN minimizes deployment costs and disruptions.
- Reliable and Efficient WSR Operation: Careful consideration is given to the management strategies employed within the WSRs, balancing the need for high reliability and performance with the constraints of an embedded system operating on harvested energy. Through intelligent resource allocation and power management techniques, we ensure the efficient and sustainable operation of the WSRs, maximizing their lifespan and contribution to the Green EMN.

The Green EMN proposed in this paper offers a compelling solution for environmental monitoring in campus environments. By combining intelligent power management, robust security measures, a scalable architecture, and efficient WSR operation, the Green EMN creates a self-powered, reliable, and secure platform for comprehensive environmental data collection and analysis. This ultimately fosters a healthier, more sustainable learning environment for students and faculty, while also contributing to broader environmental monitoring and management efforts.

**REFERENCES**.